\documentclass[preprint]{aastex}
\received{}
\revised{}
\accepted{}

\ccc{}
\cpright{none}{}

\slugcomment{}

\shorttitle{The VLT-VIRMOS Mask Manufacturing Unit}
\shortauthors{G. Conti et al.}

\begin{document}

\title{The VLT-VIRMOS Mask Manufacturing Unit}
\author{G. Conti, E. Mattaini, L. Chiappetti, D. Maccagni, E. Sant'Ambrogio, D. Bottini, B. Garilli}
\affil{CNR--Istituto di Fisica Cosmica "G. Occhialini", via Bassini 15, I-20133 Milano, Italy;
conti@ifctr.mi.cnr.it}
\author{O. Le F\`evre, M. Saisse, C. Vo\"et}
\affil{Laboratoire d'Astrophysique de Marseille, Traverse du Siphon, F-13376 Marseille, France;
Olivier.LeFevre@astrsp-mrs.fr}
\author{O. Caputi, E. Cascone, D. Mancini, G. Mancini, F. Perrotta, P. Schipani}
\affil{Osservatorio Astronomico di Capodimonte, via Moiariello 16, I-80131 Napoli, Italy;
mancini@cerere.na.astro.it}
\and
\author{G. Vettolani}
\affil{CNR--Istituto di Radioastronomia, via Gobetti 101, I-40129 Bologna, Italy; vettolani@ira.bo.cnr.it}

\begin{abstract}
The VIRMOS Consortium has the task to design and manufacture two spectrographs for ESO VLT, VIMOS (Visible
Multi-Object Spectrograph) and NIRMOS (Near Infrared Multi-Object Spectrograph). This paper describes how the
Mask Manufacturing Unit (MMU), which cuts the slit masks to be used with both instruments, meets the scientific
requirements and manages the storage and the insertion of the masks into the instrument. The components and the
software of the two main parts of the MMU, the Mask Manufacturing Machine and the Mask Handling System, are
illustrated together with the mask material and with the slit properties. Slit positioning is accurate within 15
$\mu$m, equivalent to $0.03\,\arcsec$ on the sky, while the slit edge roughness has an $rms$ on the order of 0.03
pixels on scales of a slit $5\,\arcsec$ long and of 0.01 pixels on the pixel scale ($0.205\,\arcsec$). The MMU has
been successfully installed during July/August 2000 at the Paranal Observatory and is now operational for
spectroscopic mask cutting, compliant with the requested specifications.
\end{abstract}


\section{Introduction}
The VIRMOS (Visible and InfraRed Multi-Object Spectrographs) project is the response by a French-Italian
Consortium of astronomical institutes to the ESO request for two spectrographs with enhanced survey capabilities
to be installed at the Very Large Telescope (VLT). It consists of the twin instruments VIMOS (VIsible
Multi-Object Spectrograph) and NIRMOS (Near InfraRed Multi-Object Spectrograph), with a large field of view split
into four quadrants and a high multiplexing factor in their Multi-Object Spectroscopy (MOS) observing modes
\citep{Lefevre98, Lefevre00}. VIMOS is going to be offered to the European astronomical community starting July
2001 at VLT-UT3, and NIRMOS is expected to be operational at the end of 2002 at VLT-UT4. In MOS mode, both
instruments make use of slit masks which the astronomer can design following the observational requirements.
Thus, the VIRMOS project includes the delivery to ESO of a complete, independent, off-line facility to
manufacture and handle the masks needed by the two spectrographs.\\

Multi-object spectrographs currently operational or forthcoming at other telescopes have adopted machining
solutions ranging from laser to punching to milling machines and either aluminum or carbon fiber as masks: the
MOS/OSIS instrument at CFHT (Canada France Hawaii Telescope) uses $75\,\mu$m thick black anodized aluminum masks
cut by a laser machine \citep{Dibiagio}; at the KECK telescopes the LRIS instrument uses  mechanically punched
0.4 mm thick aluminum masks \citep{Oke} while the DEIMOS
instrument\footnote{http://www.ucolick.org/~phillips/deimos/masks.html} will use 0.25 mm masks prepared by a
milling machine; the GMOS spectrograph for the Gemini telescopes uses 0.2 mm thick carbon fiber masks cut by a
laser machine \citep{Szeto}.\\

In the VIMOS case we selected the laser technique to cut the masks into 0.2 mm thick invar sheets. In this paper,
we will describe our Mask Manufacturing Unit (MMU), which is operational at the Paranal Observatory since August
2000. The MMU is presently used to provide the instrument FORS2 with the masks to be used in its Mask eXchange
Unit (MXU) \citep{Schink} and will be used with VIMOS as it arrives in Paranal in early 2001. In section 2 we
illustrate the requirements and specifications the MMU had to satisfy and in section 3 the adopted hardware
configurations. Sections 4 and 5 respectively describe the tests and tuning  of the Mask Manufacturing Machine
(MMM) and the resulting performances. Finally, Section 6 outlines the MMU operational concepts.

\section{MMU Requirements and Specifications}

\subsection{Masks}
The VIMOS (and NIRMOS) focal plane is divided into 4 quadrants, therefore 4 masks (1 mask set) are needed for
every MOS observation. The scale at the focal plane where masks must be inserted is 0.578 mm arcsec$^{-1}$ and
each quadrant has a field of view of $7\,\arcmin \times 8\,\arcmin$ ($8\,\arcmin \times 6\,\arcmin$ NIRMOS). This
defines the gross dimensions of the masks, and together with the expected best seeing at the VLT ($\sim\, 0.3\,
\arcsec$), the minimum slit width. Although the VLT Nasmyth focal plane has a curvature radius of 2 m, the VIMOS
Focal Plane Adaptation Lens brings this radius to 4 m and the tilting of the quadrants optical axes by $\sim
2\,\degr$ makes possible to use flat masks with a maximum allowed thickness of 0.4 mm. Fig. 1 shows an outline of
the VIMOS focal plane with its reference system. The size of the masks is $305\times 305$ mm, and the contour
geometry is determined by the mechanical interface with the focal plane and with the automatic mask
insertion mechanism.\\

\placefigure{fig1}

The number of slits in each mask can be more than 200, depending on slit length and spectral coverage. The
highest multiplexing factor can be attained observing with the low resolution ($R\sim 200$) VIMOS grisms, when up
to 5 rows of spectra can be stacked on each CCD. Typical slits are rectangular with width $\geq 200\,\mu$m
($0.35~\arcsec$), but the possibility to cut also curved slits must be supported.\\

The roughness of the slit edges must be as low as possible on spatial scales between the detector pixel size
(0.12 mm at the focal plane for VIMOS) and the slit length (typically 6 mm, equivalent to $10\,\arcsec$), to
optimize sky subtraction in spectroscopic data reduction. This specification has been quantified as follows: the
$rms$ roughness at pixel-size scales must be $\leq\,2\,\mu$m and the slit edge waviness must be
$\leq\,3\,\mu$m. The slit edge roughness shall be regularly inspected to verify the machine performances.\\

The mask cutting speed shall be sufficient to allow the production of the masks for 8 observing fields (32 masks)
in an 8 hour work shift (15 min per mask). For a typical 200 slit mask the total length to be cut is about 4500
mm (including the overhead for the mechanical interface) and therefore the requirement translates in a cutting
speed $\geq 6$ mm s$^{-1}$.\\

One of the most critical items is the slit positioning accuracy in the focal plane, which eventually determines
whether the chosen set of objects is going to be found in the slits when pointing the telescope. The error budget
is made of several components: (1) the mapping accuracy of the CCD into the focal plane, i.e. the accuracy with
which optical distortions are taken into account; (2) the mask scale variations, i.e. the thermal expansion of
the mask material due to the temperature variations between mask manufacturing and spectroscopic observation; (3)
the positioning accuracy of the slit cutting machine; (4) the accuracy of the positioning of the mask in the
focal plane and its repeatibility. The overall acceptable positioning accuracy is 1/3 of a pixel (slits of
$1\,\arcsec$ width project onto 5 pixels), which means $\leq 30\,\mu$m. Component (1) contributes for 3 $\mu$m,
the $rms$ of the fit of the transformation matrix from pixels to mm in the focal plane; component (4) is
estimated to be $\sim 10\,\mu$m, since the mask contours can be customized to the as-built focal plane mechanical
interface. Thus, $\sim 17\,\mu$m are allowed for component (2) and (3).\\

The mask surfaces shall have the lowest possible reflectivity at the operating wavelengths (370 to 1850 nm).

\subsection{Mask Handling}
To minimize observing overheads, and to avoid people inside the domes during the night, MOS instruments at the VLT
must be able to change the mask configuration through remote control: in our case, both VIMOS and NIRMOS can have
15 different mask sets at any time. For each instrument and for each quadrant there is a demountable Instrument
Cabinet (IC) that must be prepared by the MMU operator by inserting the requested masks as much in advance of the
spectroscopic observations as possible. These requirements imply that a large number of masks is kept available
at any time, from which to choose the ones to be inserted in the ICs for the following night(s) of observation.
Storage for 100 mask sets (400 masks) has to be provided. The masks must thus be uniquely identified, must be
traceable and their position in each IC must be certain and made known to instrument control software. A mask
handling system, providing hardware and software tools to reach the goal of having the right mask in the focal
plane at the right time, must be implemented.\\

\section{Adopted solutions and MMU configuration}

\subsection{Short history}
In the initial concept (1997), the MMM was intended as a milling machine which would cut the slits in a 0.1 mm
thin brass sheet, supported by a 10 mm thick aluminum frame. We assembled a small milling machine and we proved
that the roughness and speed requirements were fulfilled. The minimum obtainable slit width was $300\,\mu$m. This
solution was discarded because the ICs were too large and the accuracy in the slit positioning was hampered by
the thermal expansion of brass, given the possible temperature differences between the time a mask was
manufactured and used in the instrument or the temperature variations during the observations. Furthermore the
lifetime of the cutting tools was short due to breaking and to wear (that caused a slit width variation).
Subsequent developments (1998) were aimed at minimising the sources of errors, by making use of aluminum and
thicker (but still  $<\,0.3$ mm), frameless masks, which had the advantage of reducing the size and the weight of
the ICs. The next natural step was the use of a material with a low thermal expansion coefficient: we tested
carbon fiber, kevlar, graphite and invar, but it was impossible to obtain the required slit edge quality with a
milling machine. Only recently a new type of laser cutting machines, called Stencil Laser, became available on
the market. In particular, one machine could meet the specifications by making use of 0.2 mm thick invar sheets.
The cutting speed makes possible to cut the mask contour on the machine itself, thus the masks can be frameless
and customised to the quadrant interface where will have to be mounted, and, as a further bonus, also any slit
width $\geq\,80\,\mu$m became a possibility. For a full account of the trade-off between the two solutions
(milling and laser cutting) and for the choice of the laser machine manufacturer, see \citet{Conti}.

\subsection{The mask manufacturing machine}

\placetable{tbl-1}

Our choice for the MMM has been the StencilLaser 600 by LPKF whose main characteristics are given in Table~1. The
machine uses a pulsed Nd:YAG laser head, that, for normal operation, is classified for safety as a class~1 laser
product (as safe as a compact disk player). The laser head is fixed while the working platform, on which the
invar sheet is mounted, is moved in XY by two servo-motors over 4 air pads. Cutting occurs under 16 bars
compressed air jet. The whole cutting process is controlled by the LPKF StencilMaster software running on a PC
called Mask Manufacturing Control Unit (MMCU). A picture of the machine can be seen in Fig. 2.\\

\placefigure{fig2}

\subsection{The mask material and the raw mask preparation}
The chosen mask  material is invar, a 36\%Ni-64\%Fe alloy\footnote{by Krupp VDM, Werdohl, Germany, trade name
Pernifer 36} with a thickness of 0.2 mm. The main mechanical and thermal characteristic of invar compared with
aluminum and stainless steel are given in Table 2.\\

\placetable{tbl-2}

Mask manufacturing includes, as a last operation, the cutting of the contour, therefore the raw invar sheets
shall have larger dimensions than the masks, to fix them on the working platform of the laser cutting
machine. We adopted $450\times 340$ mm sheets, that weigh 0.250 kg each.\\

 \placefigure{fig3}

The yearly mask need for VIMOS is estimated to be $\sim 2000$. One aim of our work was to find a method to
prepare a large quantity of raw masks at the lowest cost. Invar is delivered in rolls with the requested width
and must be coated with a black anti-reflection paint. The requested characteristics for the coating are:
thickness $<\,20\, \mu$m; good adhesion to the metallic substrate; dull black color; uniformity of the coating
over the 2 surfaces. The adopted procedure\footnote{implemented by CEM Lavorazioni Elettrochimiche srl, Milano,
Italy} for the raw mask preparation consists of: chemical and mechanical cleaning; black coating of the two sides
of the strip, using a roller system; warm curing of the varnish; insertion of a low adhesion plastic protective
film; straightening and de-stressing to eliminate the roll curvature; cutting to the requested dimensions; visual
inspection for absence of wrinkles and scratches; packing in folders of 20 sheets each. This industrial procedure
has the disadvantage that the first and the last part of the strip must be discarded because mechanically
damaged. For a 450 mm wide strip this loss can be quantified in about 100 kg of material. However, we obtained
4400 sheets out of 1200 kg of invar, at a cost of $\sim 11$ DM per sheet. Fig. 3 shows a typical VIMOS mask.\\

Measurements of the specular and diffuse reflectance of coated invar samples at 633 and 1150 nm give the results
shown in Table 3.\\

\placetable{tbl-3}

\subsection{The Mask Handling System}
The concept of the Mask Handling System (MHS) is to provide the hardware and software necessary to control the
mask flow.\\

Mask identification is done by direct cutting of a 6 digits bar code at 3 mm from the edge of the masks using the
European standard 2/5 Interleaved Code (EN 801). The most significant digit of the code identifies the quadrant
(from 1 to 4 for VIMOS and from 5 to 8 for NIRMOS). Two bar code scanners\footnote{Datalogic, Bologna, Italy,
mod. DS2100-1000} provide the decoding during all the operations of the MHS.\\

All the masks that, at a given time, are not inserted into the ICs are stored in the Storage Cabinet (SC) of the
relative instrument.

Fig. 4 shows a picture of the in house manufactured VIMOS SC. It can store 400 masks (100 mask sets) inserted
vertically into transparent polycarbonate slots in such a way that the bar code can be read by the scanner
mounted on a sliding stage. Because of space constraint, the VIMOS SC has 2 quadrants per side and the bar code
scanner must be moved from one side to the other with the connection cable suspended on the ceiling by a
rotating pivot. \\

\placefigure{fig4}
\placefigure{fig5}

The 15 masks in each Instrument Cabinet (which is part of the instrument and not of the MMU), are packed with a
pitch of 4.5 mm. Before observations the ICs must be prepared by unloading the masks no longer necessary and by
loading the requested new ones. This operation is quite delicate because a mask in a wrong IC or in a wrong slot
in the IC will jeopardise the observation. Mask insertion is thus controlled by a robot\footnote{produced by
ANTIL srl, S. Giuliano Milanese, Italy}, shown in Fig. 5, that performs a software controlled Z--displacement.
The 4 ICs are mounted in fixed order (constrained by the mechanical interfaces) into an IC box that is placed on
the robot platform. The robot moves it with an accuracy of about 0.1 mm, in order to place the appropriate IC
slot in front of a mask stand (see Fig. 5). The Z positions of the slots are calibrated and stored in a table
used by software to command the robot displacement. Then the unloading/loading operation is done manually by
means of a moving clamp, with the assistance of a bar code scanner identical to the one used in the SC. This bar
code scanner is also used for off-line mask identification (storage after manufacturing).\\

The third part of the MHS is the Mask Handling Control Unit (MHCU), a PC that runs the Mask Handling Software
(MHSw), the LPKF CircuitCam software and connects the bar code scanners and the IC Robot.

\subsection{Software components}

\placefigure{fig6}

The MMU software components developed in house to support the operational procedures described in section 6 are:
\begin{itemize}
\item The MHSw running on MHCU which provides a Graphical User Interface (GUI) (see Fig. 6) for all mask
handling functions except the actual manufacturing, records where masks are stored, handles the communication
with Instrument/VLT software and acts as a front end to the LPKF CircuitCam software (see 6.1)
\item Cut Manager software running on MMCU which is a simple GUI which assists orderly usage (and archiving)
of mask files and acts as a front end to LPKF StencilMaster, used to cut the masks
\item a Mirroring Module that runs on both computers to keep synchronised the various data files on the disks,
allowing easy recovery in case of failure of one.
\end{itemize}
All software is developed using Microsoft Visual Basic 6.0 under Windows NT operating system.

\subsection{The quality control equipment}
The cutting performance of the laser machine from the point of view of the slit edge roughness must be
periodically controlled to be inside the specifications reported in 2.1. We have defined the following quality
control procedure: 9 square samples (30 mm side) are cut from an invar sheet under the same laser machine
settings as the mask slits. The roughness of the sample borders is measured by a mechanical roughness
meter.\footnote{Taylor Hobson Ltd, Leicester, UK, mod. Talysurf Plus} The samples are placed vertically under the
measuring probe, which is a 1 mm wide chisel edge stylus with $5\,\mu$m tip radius, that measures the roughness
over the full thickness of the 0.2 mm samples. The roughness can be due to both the laser cutting process itself
and the residual invar melting flashes. The instrument is connected to a third PC where all measurements carried
out during the lifetime of the MMU can be acquired, analyzed and archived. The slit width is checked by a
microscope with a calibrated reticle.

\section{Test and tuning of the laser cutting machine}

\subsection{Performance verification}
The StencilMaster software by LPKF controls the whole cutting process, i.e. the displacement of the working
platform, and the laser parameter settings and switching. The XY motion system is controlled by rotary encoders on
the motor shafts, but a calibration of the system itself can be done using the linear glass rules provided on the
two axes. During such procedure the working platform moves, backwards and forwards, along the Y and X direction,
with a predefined step to cover the whole traveling range. After each step the displacement is read on the rules
and the difference between the requested and the actual positions is stored in a calibration file. During normal
cutting operation such correction is applied to the requested position to take into account this difference. A new
calibration of the motion system is recommended if a significant temperature variation occurres or if the machine
is switched off for some days. A typical calibration file is plotted in Fig. 7.\\

\placefigure{fig7}

To verify the accuracy of the working platform displacements after calibration, we used a laser interferometer
system\footnote{Hewlett Packard mod.5526A} with a resolution of $0.1\, \mu$m and an accuracy of $\pm\,0.5\, ppm$.
Normally the readings from internal linear glass rules are not accessible while the machine is moving. For this
reason a single (uncalibrated) step displacement was commanded manually and, when done, the internal rule reading
was recorded together with the interferometer displacement measurement. The agreement between the two measures was
within $\pm\, 2 \,\mu$m, that confirms the accuracy of the LPKF machine displacements. We did further tests
asking for selected linear cuts along the interferometer beam (alternately mounted along X and Y) and reading the
interferometer readout and the glass rules at the starting and ending positions. We have verified that, during
cutting, the movements are commanded in calibrated units, with the expected accuracy.\\

The laser beam correctly focalised has a diameter of $40\, \mu$m. The CircuitCam program that calculates the
cutting paths, takes into account this value to obtain the requested slit dimensions. The slit width measurements
show an error of $\pm\,2\, \mu$m, independently of the slit width.

\subsection{Cutting tool tuning}
The file describing each mask (see 6.1) contains the mask slit pattern, the bar code, the other features and the
appropriate quadrant contour, associated to different "layers". To each layer corresponds a cutting tool, i.e. a
particular cutting speed and laser set-up parameters. Among several possible choices, we carried out a series of
tests to find the best tool (called "cutting fine") which would satisfy the edge roughness specifications while
working at the necessary speed, while a faster but rougher "cutting" tool is used for contour and lower precision
features. Table 4 gives the adopted tool settings.\\

\placetable{tbl-4}

The speed and frequency are fixed parameters while the others are adjusted at each laser lamp change to keep the
power constant. The best focus of the laser beam must be positioned inside the thickness of the cutting material.
It can be adjusted by a micrometer that determines the distance of the focusing lens from the sheet during
cutting; we have verified that also this parameter influences the edge roughness and we have optimized it (this
tuning has been repeated after the installation in Paranal).\\

\section{Resulting Performances}

\subsection{Mask cutting time}
The time needed for mask manufacturing includes a fixed overhead of about 460 s for invar sheet mounting
operations, connection set-up, bar code and contour cutting, and scales up to 950 s for a mask with
200 slits. Thus, at most two 8 hours shifts are necessary to cut the 60 masks which would fill the 4 ICs
of one instrument.\\

\subsection{Slit positioning accuracy}
Having made allowance for the accuracy of the CCD mapping onto the focal plane and for the repeatability of the
mask positioning in the focal plane (see 2.1), we must verify that the invar thermal expansion and the
positioning accuracy of the laser machine allow us to place slits within $17 \mu$m. Mask manufacturing
occurs at a controlled temperature of $20^{\circ}$C while the average temperature at Paranal is $10^{\circ}$C,
thus the (negative) invar expansion over the mask size is only $\sim 2 \mu$m.\\

To verify the real accuracy of the slit positioning we cut several masks with a pattern of square holes 3 mm wide
and with the usual contour. We used the LPKF machine itself to measure the aperture positions, by mounting a
microscope fixed to the laser head. The mask was laid on an invar sheet secured to the working platform and,
after a preliminary alignment, the procedure was to move it in small steps until the hole edges were centered
under the eyepiece reticle and to read the displacements using the internal glass rules. When a mask has to be
cut, the invar sheet is fixed on the working platform by means of manual clamps and two pneumatic pistons apply
an adjustable force to keep it flat. The results obtained from the measures show that the accuracy of the aperture
positioning on cut masks depends mainly on the tension applied during cutting: a high force (50 kg) means an
elastic elongation of the sheet that is released when the contour is cut and smaller distances with respect to
the nominal ones are systematically measured. A low force ($<10$ kg) is not sufficient to straighten the sheet
and produces higher distance values. At the end of an optimization process, which required a long series of
measurements, we adopted a tensioning force of $\sim 18$ kg as the best trade-off. To have a more repeatable and
stable tensioning force we have substituted the pneumatic pistons with 9 mechanical springs, applying 2 kg each.
The final evaluation is that a positioning accuracy $\le\,15\,\mu$m is verified.\\

\subsection{Slit edge roughness}

The fine tuning of the laser cutting machine was carried out having the slit edge roughness as driving parameter.
A protocol was thus established implying the cutting of a number of samples suitable for roughness measurements,
performed using the Talysurf instrument described in 3.6. The sample sides are scanned one at a time on its
thickness by the chisel edge stylus and the roughness profiles are acquired and stored for off-line analysis.
First the slope introduced by the sample non-horizontal mounting is removed, by a linear fit, and then the
rectified profile is successively filtered by two high pass gaussian filters. First, with a cut-off length of
0.12 mm (1 pixel of the CCD), the parameter $W_{q}$, that is the $rms$ of the roughness at pixel-size scales, is
computed from the filtered profile, over a scan length of 25 mm. Second, with a cut-off length of 2.5 mm, to
calculate the parameter $W_{t}$, that is the maximum peak to valley of the edge waviness. The samples prepared for
roughness measurements must be handled and mounted under the the probe very carefully because any mechanical
shock makes the sample unusable; we discovered that dust and any dirt deposited on the sample border can easily
mimic a bad cutting performance: thus samples measurements representative of the slit edge roughness which will
be found in the masks, must be carried out in the same environmental conditions as mask manufacturing, i.e. in an
air conditioned, reasonably clean room. The quality control protocol was applied before and after the shipment of
the MMU to Paranal. The roughness measurements obtained on 9 samples in Europe and in Chile showed no significant
difference. The obtained mean values are shown in Table 5 and a typical roughness meter plot is shown in Fig. 8.\\

\placetable{tbl-5}

\placefigure{fig8}

\subsection{Scientific implications}
The results obtained not only meet the specifications, but, even more important, guarantee that observations
will yield good quality data. Thanks to the slit positioning accuracy we obtained through the choice of
the mask material and the fine tuning of the cutting operations, objects on the sky can be placed in slits
all over the unvignetted Nasmyth focus covered by VIMOS, making really possible to obtain several hundreds
of meaningful spectra of faint objects in one exposure. The excellent slit edge roughness quality guarantees
optimum sky subtraction, especially of bright sky lines regions, even in case of low signal-to-noise.

\section{MMU operation concepts}
The MMU operation concepts have been devised to provide a smooth and orderly mask flow. At least for some time
since after VIMOS commissioning, pre-imaging is a requirement to be able to design masks. Once the files describing
a mask set have been prepared by the astronomer, they must reach the MMU so that mask production can start.
Since the spectroscopic observations can be scheduled up to a few months later, masks must be identified and stored
so that they can be univocally retrieved when they must be inserted in the ICs just before the observations.
The instrument control software must be made aware of the IC position in which a specific mask is to be found.
Furthermore, the SCs have a limited capacity and from time to time it is necessary to discard masks which have
a low probability of being used in order to leave room for new masks.\\

MMU routine operations will be totally slaved to orders issued by Instrument software modules running on separate
workstations. The only module actually talking to MMU is, in the case of VIRMOS, an instrument specific Mask
Conversion Software (MCS module) whose purpose, besides transmitting Job orders and retrieving Termination
reports for all activities, is also, for the mask manufacturing case, to translate the slit positions on the sky
(selected by the astronomer using the Mask Preparation Software, see \citet{Garilli}) from pixel to mm with
respect to the quadrant optical axis using the CCD to mask transformation matrix.\\

The purpose of the entire MMU system is to perform the following operations:
\begin{itemize}
 \item to manufacture the masks on the laser machine requested via a Mask Manufacturing Job order (MMJ), and to
 store the manufactured masks in the SCs
 \item to load the masks requested via a Mask Insert Job order (MIJ) into the ICs
 \item to get rid of masks and related files no longer needed as requested via a Mask Discard Job order (MDJ)
 \item to manage and recover contingency conditions which may occur during the above tasks.
\end{itemize}
In all operations the MMU will only know masks by their individual six-digit code, and will normally handle
entire mask sets. A mask set is defined as a set of four masks, with the same last 5 digits in the id, used for
the same observation in the 4 VIMOS or NIRMOS quadrants. All operations (except contingency recovery and
maintenance operations managed autonomously by the MMU operator) are executed after receipt from MCS of a Job
order in the form of a file transferred into a staging area of the MHCU, and terminate upon placing a Termination
report in the same staging area, to be retrieved by MCS that has also access to the tables describing the current
content of the SC (SCT) and of the ICs (ICT). The SCT, for each quadrant, is a list of 100 items containing the
codes of the masks currently stored or "empty", with no information about the slot where the mask are located.
The ICT, for each Instrument Cabinet, is a list of 15 items which combine the slot number with the code of the
contained mask.

\placefigure{fig9}

\subsection{Mask manufacturing and storing}
The information about the slit positions and sizes is contained in the Machine Slit Files (MSF) transmitted
together with a MMJ. The full manufacturing cycle is a 4--step operation (Fig. 9).
\begin{enumerate}
 \item The MHSw function {\tt convert} selects an MMJ and
  converts the MSF files into the Gerber \citep{Gerber} CAD format, accepted in input by the
  LPFK CircuitCam software. The MSF to Gerber conversion takes care of items relevant for masks,
  like the correction for the calibrated offsets of the optical axis with respect to the mask centre, and the
  insertion of the mask contour, fixing holes and other control apertures (see Fig. 3) which are
  merged from template files and assigned to appropriate layers. This arrangement is highly modular and
  allows to tune and replace such characteristics without recompiling the code, while at the same time keeps
  effectively decoupled the parameters specific of mask manufacturing from those specific of the astronomical
  observation.
 \item  LPKF CircuitCam software converts the Gerber files into proprietary format ({\tt lmd})
  files, transferred to MMCU and accepted by the StencilMaster software, at the same time optimizing
  the cutting path for the laser beam, and associating the layers with the cutting tools and manufacturing phases.
 \item The operator places an invar sheet on the working platform of the laser machine, and runs the function Cut
  Manager on the MMCU, which gives him access to StencilMaster to cut the mask.
 \item At the same time, the operator runs the MHSw function {\tt store} on the MHCU, which, by reading the bar
  code, indicates in which quadrant section of the SC the mask must be placed. Acknowledgement of these operations
  will automatically update the SC table and write the MMT (Mask Manufacturing Termination report) which will
  eventually be retrieved by the MCS.
\end{enumerate}

\subsection{Instrument cabinets preparation}
When the ICs must be loaded with a new set of masks, the operator finds a MIJ on the MHCU. The ICs coming from the
telescope are inserted in the IC box which is placed on the robot and the operator runs the {\tt IC preparation}
function of the MHSw. Loading the ICs is a 2 step process: this is accomplished by running first the {\tt unload}
(sub)function, at the end of which the no-longer needed masks are returned to the SC while the masks still needed
are left in place; the IC and the SC tables are updated. Then the {\tt load} (sub)function is started, which
allows to search the requested mask in the SC and to insert the masks in free IC slots, by appropriately moving
the IC box. The search in the SC occurs manually, moving the bar code scanner until the requested mask is located
(computer gives an audio signal as long as the scanner beam is located in front of the wished mask). The bar code
is double checked by the scanner located on the robot stand before inserting the mask into the IC. The SC and IC
tables are updated again and a report (MIT) is made available to MCS. The report and the updated tables are
propagated to be used for mask insertion management at the instrument focal plane.

\subsection{Mask discarding}
To make room in the SC for newly manufactured masks, a MDJ must be issued whenever a mask set is no longer
required (either the observation has been executed or expired because of celestial constraints). The {\tt discard}
function should normally have priority with respect to any other. It allows the operator to search (as described
in 6.2) for the masks to be discarded and to remove them from the SC, thus deleting the files of the mask from the
archives, updating the SC table and issuing the final report.

\subsection{Recovery functions and error handling}
Great care has been dedicated to manage the possible errors during mask handling (e.g. mask dropped or damaged
during manual displacements, wrong insertion into the SCs or ICs, problems in bar code scanner or robot connection
etc.). All the error messages are listed in the user's manual and  contain the suggested recovery procedure using
the 8 provided recovery functions.

\section{Summary}
The VIRMOS Mask Manufacturing Unit is now operational at Paranal Observatory, following verification that
the specifications set at the beginning of the project have been met. The laser cutting machine has been
finely tuned with an optimal choice of the parameters, and, although the MMU is largely assembled from
industrial products, every effort has been made to design and implement software tools which would make the
mask flow nicely fit in the VLT data flow system. The slit quality that the VIRMOS MMU provides makes it
the best operational tool for multi-object spectroscopy applications.

\acknowledgements
The VIRMOS-MMU manufacturing is done under ESO contract 50979/INS/97/7569/GWI. The MMU development was aided by
financial support from Italian CNR and CNAA. We warmly thank LPKF technical staff (T. Nagel, A. M\"uller and S.
B\"onigk) for their appreciated collaboration during the tuning of the laser cutting machine and N. D'Addea
(CNR-ITIA) for lending us the laser interferometer system. We also thank G. Avila, and the Paranal ESO
Engineering staff for their help during installation of the MMU at the Observatory.

\clearpage

\figcaption[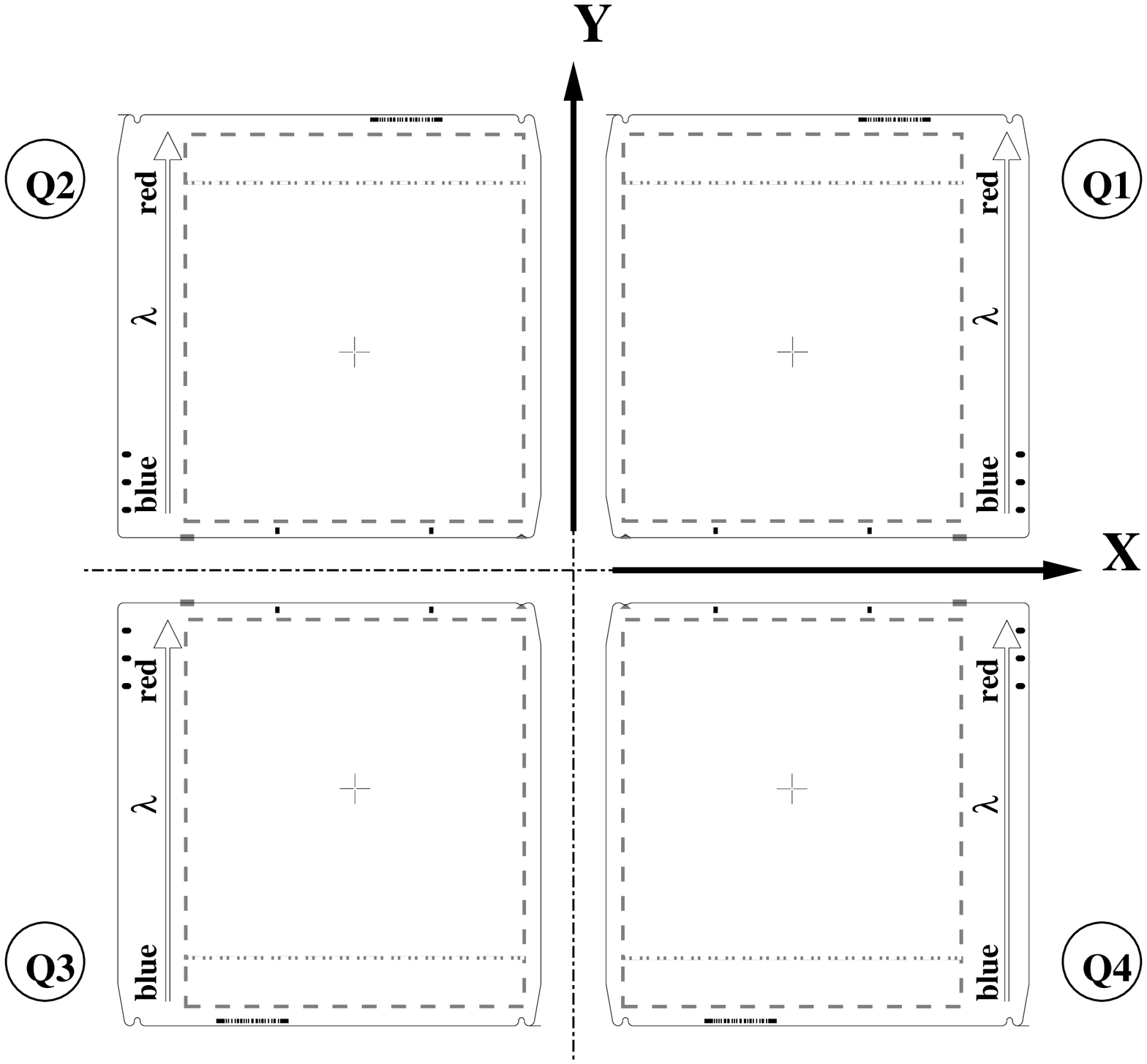]{The figure shows (on scale) the location in the 4 VIMOS quadrants of the 4 masks
     with all contour and other details (see Fig. 3) and respective orientation in the common reference system (which
     is oriented as the MMM reference system). The dispersion direction of the spectra is also indicated. The cross
     marks the position of the optical axis (which is not coincident with the mask centre), while the dashed line
     indicates the $244\times 279$ mm "useful area" corresponding to the detector field of view. The dotted line
     delimits a square $244\times 244$ mm area around the optical axis. \label{fig1}}

\figcaption[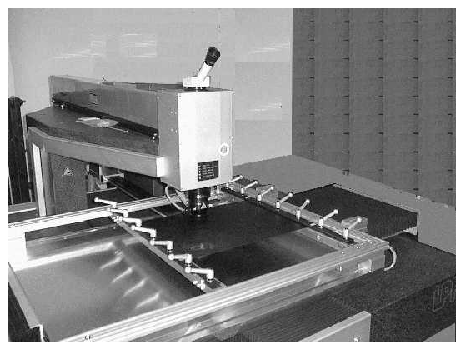]{The LPKF laser cutting machine installed in Milano.
     An invar sheet is mounted and clamped on the working platform. For mask cutting the laser focusing system
     moves down close to the sheet and the whole platform is displaced in the X and Y directions. \label{fig2}}

\figcaption[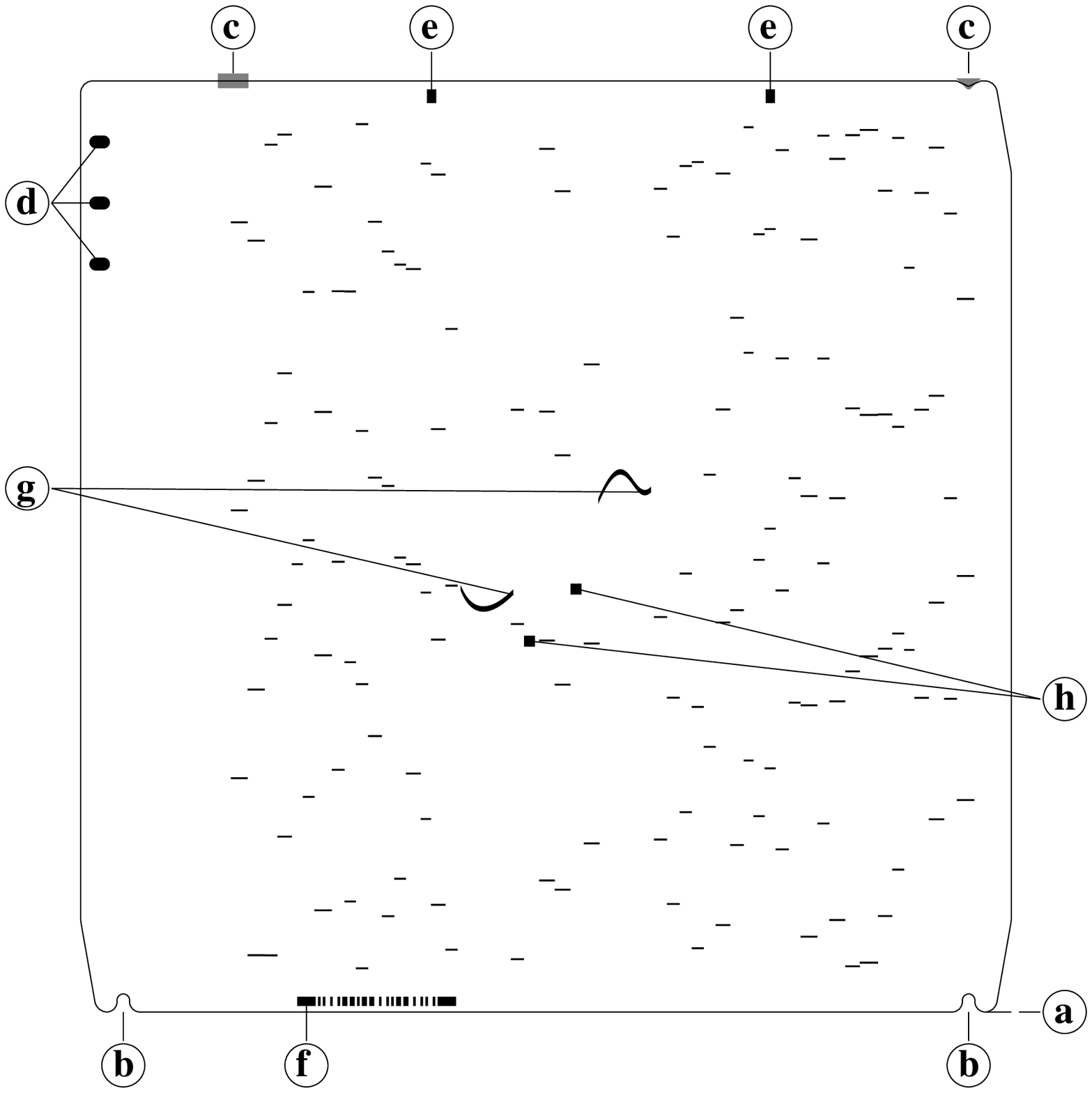]{Layout of typical VIMOS mask (quadrant 3) showing contour
     and slits and the following features: (a) the cutting of the contour starts outside the mask area to have a
     clean outline; (b) the two bottom notches are used to position the mask inside the IC; (c) the two upper
     notches (in gray), a rectangular and a triangular one, make the reference for mask positioning in the focal
     plane: they are cut with the same accuracy as the slits; (d) the three oval holes are the grabbing points of
     the clamp moving the mask from the IC to the focal plane and back; (e) the two rectangular fixing holes are
     used to hold the mask in position in focal plane; (f) the 6-digit bar code used to identify the mask; (g) two
     curved slits; (h) two square slits corresponding to the positions of two reference objects used for fine
     alignment of the mask on the sky. \label{fig3}}

\figcaption[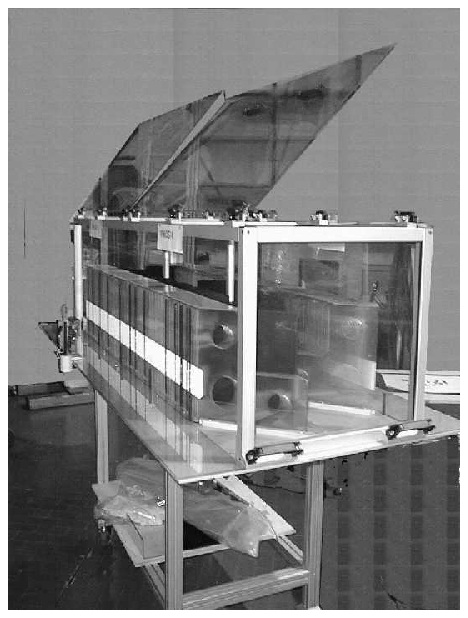]{The VIMOS Storage Cabinet. The white surfaces are used as background for the bar code
     during mask identification. \label{fig4}}

\figcaption[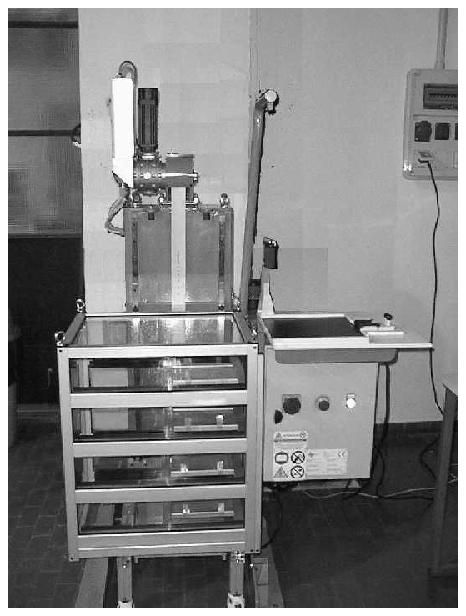]{The VIMOS Instrument Cabinet Robot with the empty IC box placed on its platform and
     a mask positioned on the stand. \label{fig5}}

\figcaption[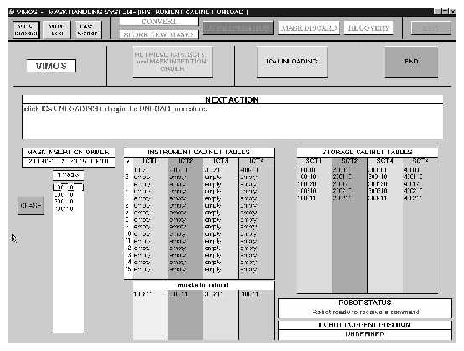]{Example of a GUI of the MHSw ({\tt IC preparation/unload} (sub)function)
     \label{fig6}}

\figcaption[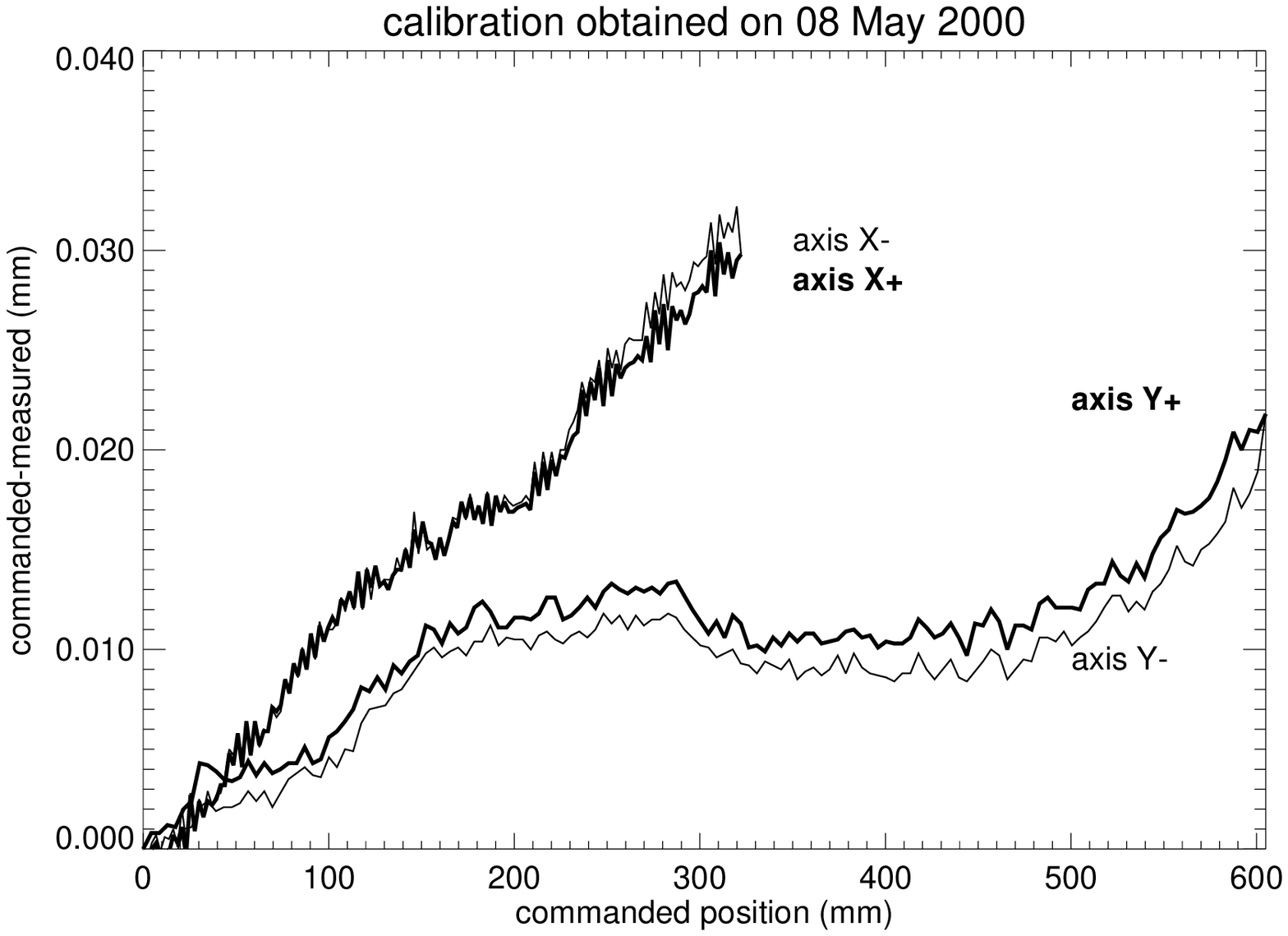]{The content of a typical laser machine
    calibration file. The abscissa of the plot correspond to a commanded excursion on the X or Y axis of the
    machine (for our application we use the whole X excursion but only the central part of the Y excursion).
    The calibration procedure records the difference between the requested position
    and the actual position measured by the linear glass rules (this difference is shown in ordinate) once
    moving in the forward (positive) direction (bold line) and once in the backward (negative) direction (thin
    line). The difference between the two curves gives an indication of the backlash of the motion system. Actual
    movements during cutting are compensated using the content of the calibration file (by default the positive
    direction curve is used). \label{fig7}}

 \figcaption[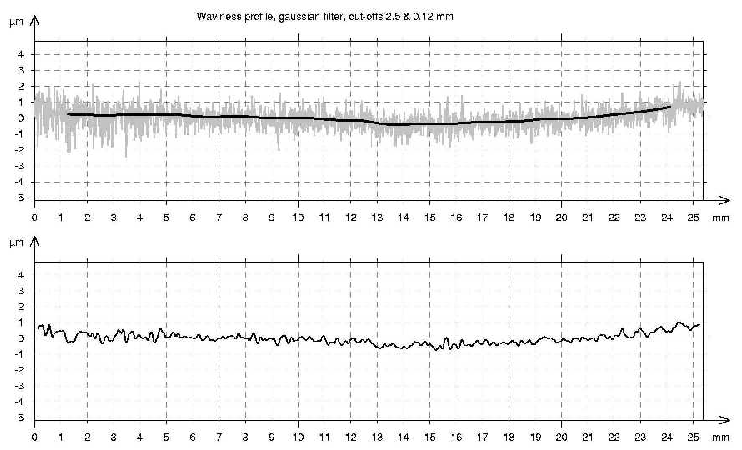]{Typical roughness meter plots of a standard measurement sample. The
     linearized rough data are shown in gray in the top panel together with the filtered (waviness) profile
     using a 2.5 mm cutoff (bold line). This particular sample has a $W_{t}$ of $1.13\,\mu$m. The bottom panel
     contains the profile filtered with a 0.12 mm cutoff, which gives the pixel-scale roughness. This
     particular sample has a $W_{q}$ of $0.337\,\mu$m. \label{fig8}}

\figcaption[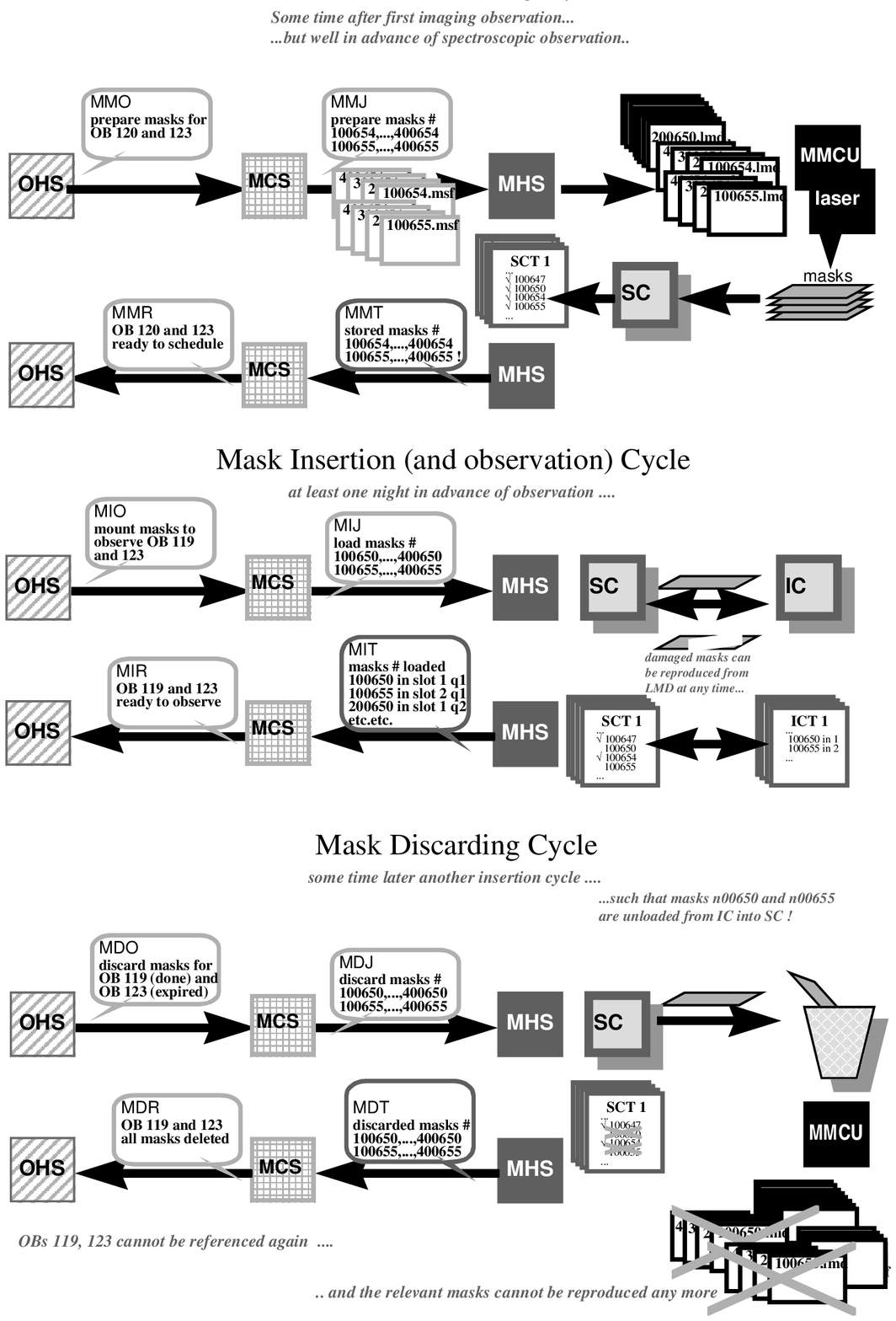]{Typical mask handling flow. Hatched boxes indicate VLT/Instrument
     software modules (like OHS Observation Handling Software, and MCS Mask Conversion Software), while filled
     boxes indicate MMU components. In the mask manufacturing cycle, from Mask Manufacturing Orders (MMO,
     arranged by Observing Blocks, OBs), MCS generates Mask Manufacturing Jobs (MMJ) issued to MMU, which
     processes the associated files (.msf and .lmd), produces masks, stores them in the SC, updates the SC Table
     (SCT) and prepares the MMT report from which MCS generates the final Mask Manufacturing Report (MMR). In the
     mask insertion (and observation) cycle, from Mask Insertion Orders (MIO), MCS issues MIJs to MMU, which
     exchanges the needed mask between SC and IC, updates the SC and IC Tables and prepares the Mask Insertion
     Termination report (MIT) from which MCS generates the final Mask Insertion Report (MIR). A mask can be
     discarded only once it has been used for observation or declared invalid by OHS and if it is in the SC. From
     Mask Discard Orders (MDO, arranged by OBs), MCS issues MDJs to MMU, which allows to search the masks to be
     discarded, gets rid of any associated file (so that the mask cannot be reproduced any more), updates the SC
     Table and prepares the Mask Discard Termination report (MDT) from which MCS generates the final Mask Discard
     Report (MDR). \label{fig9}}

\clearpage

\begin {deluxetable}{ll}
\tablewidth{0pt} \tablecaption{Laser cutting machine characteristics \label{tbl-1}} \scriptsize
\tablehead{\colhead{Item} &\colhead{Description} } \startdata
Manufacturer & LPKF Laser \& Electronics AG, Garbsen (Germany) \\
Model & StencilLaser System 600/600 \\
Stand & natural granite \\
Cutting area & $600\times 600$ mm \\
XY motion & DC servo-motors + rotary encoders, res. $0.5\,\mu$m \\
XY motion calibration & glass linear rules, res. $0.5\,\mu$m \\
Working platform & supported by four air pads on the granite stand \\
XY accuracy on full travel & $\pm 10\,\mu$m \\
Repeat accuracy & $3\,\mu$m \\
XY squareness& $<5\,\arcsec$ \\
Laser cut location accuracy & $\leq 15\,\mu$m \\
Laser cut dimensions repeatability  & $\pm\, 2\,\mu$m \\
Cutting speed range & 1-50 mm s$^{-1}$ \\
Laser head & Flashlamp pumped Nd:YAG (1064 nm), max power 60 W \\
& lamp life $\sim 400$ h, max pulse rate 4000 Hz\\
Laser beam diameter & $40\,\mu$m \\
Gas for cutting assist & air 16 bar \\
Water cooling & external chiller \\
Gas exhausting system & yes \\
Realignment after lamps changing & no \\
Laser safety (normal operation) & class 1 laser product \\
Control & PC via RS232 \\
Overall dimensions & $1750\times 2300\times 1350$ mm \\
Dimensions of control rack & $600\times 950\times 1900$ mm \\
Total weight & $\sim\,3000$ kg \\
Software & CircuitCam to generate proprietary format files \\
& StencilMaster to control the cutting operations.\\
\enddata
\end{deluxetable}

\clearpage

\begin{deluxetable}{lccc}
\tablewidth{0pt} \tablecaption{Comparison of invar characteristics \label{tbl-2}} \tablehead{
\colhead{Characteristic} & \colhead{invar} & \colhead{Al} & \colhead{steel} } \startdata
Coeff. therm. exp.($\times 10^{-6}/^{\circ}$C) & 0.8 & 23 & 9.5 \\
Mod. of elasticity ($\times 10^{9}$ N m$^{-2}$) & 143 & 69 & 200 \\
Density ($\times 10^{3}$ kg m$^{-3}$) & 8.1 & 2.7 & 8 \\
Specific stiffness ($\times 10^{6}$ m$^{2}$ s$^{-2}$) & 18 & 26 & 25 \\
\enddata
\end{deluxetable}

\clearpage

\begin{deluxetable}{lcc}
\tablewidth{0pt} \tablecaption{Coated invar reflectance \label{tbl-3}} \tablehead{ \colhead{wavelength} &
\colhead{specular reflectance (\%)} & \colhead{diffuse reflectance (\%)} } \startdata
633 nm & 0.04 & 4.7 \\
1150 nm & 0.04 & 3.7 \\
\enddata
\end{deluxetable}

\clearpage

\begin{deluxetable}{lcc}
\tablewidth{0pt} \tablecaption{Laser machine tool settings \label{tbl-4}} \tablehead{ \colhead{} &
\colhead{"cutting fine} & \colhead{"cutting"} } \startdata
Cutting speed & 6 mm s$^{-1}$ & 20 mm s$^{-1}$ \\
Laser power & 19 W & 21 W \\
Voltage & 235 V & 269 V \\
Laser pulse frequency & 1200 Hz & 1600 Hz \\
Laser pulse width & 0.19 ms & 0.11 ms \\
\enddata
\end{deluxetable}

\clearpage

\begin{deluxetable}{lcc}
\tablewidth{0pt} \tablecaption{Slit edge roughness \label{tbl-5}} \tablehead{ \colhead{Parameter} &
\colhead{measured} & \colhead{required} } \startdata
$W_{q}$ (cut-off=0.12 mm) & $0.35\,\pm\,0.08\,\mu$m & $\leq\,2\,\mu$m \\
$W_{t}$ (cut-off=2.5 mm) & $1.06\,\pm\,0.43\,\mu$m & $\leq\,3\,\mu$m \\
\enddata
\end{deluxetable}

\clearpage

\begin{figure}
 \epsscale{0.5}
 \plotone{mmu_fig1.eps}
\end{figure}

\clearpage

\begin{figure}
 \epsscale{0.5}
 \plotone{mmu_fig2.eps}
\end{figure}

\clearpage

\begin{figure}
 \epsscale{0.5}
 \plotone{mmu_fig3.eps}
\end{figure}

\clearpage

\begin{figure}
 \epsscale{0.5}
 \plotone{mmu_fig4.eps}
\end{figure}

\clearpage

\begin{figure}
 \epsscale{0.5}
 \plotone{mmu_fig5.eps}
\end{figure}

\clearpage

\begin{figure}
 \epsscale{1}
 \plotone{mmu_fig6.eps}
\end{figure}

\clearpage

\begin{figure}
 \epsscale{0.5}
 \plotone{mmu_fig7.eps}
\end{figure}

\clearpage

\begin{figure}
 \epsscale{1}
 \plotone{mmu_fig8.eps}
\end{figure}

\clearpage

\begin{figure}
 \epsscale{0.8}
 \plotone{mmu_fig9.eps}
\end{figure}
\clearpage

\end{document}